\documentclass[aps]{revtex4}
\usepackage{graphicx}

\newcommand{\ax}{{\bf a}_1}
\newcommand{\ay}{{\bf a}_2}
\newcommand{\Ax}{{\bf A}_1}
\newcommand{\Ay}{{\bf A}_2}

\begin{document}
\title{Packing Squares in a Torus}

\author{D. W. Blair}
\email[]{dwblair@physics.umass.edu}
\affiliation{Department of Physics,
University of Massachusetts,
Amherst, MA 01003-3720, USA}

\author{J. Machta}
\email[]{machta@physics.umass.edu}
\affiliation{Department of Physics,
University of Massachusetts,
Amherst, MA 01003-3720, USA}
\affiliation{Santa Fe Institute, 1399 Hyde Park Rd, Santa Fe, NM 87501, USA}

\author{C. Santangelo}
\email[]{csantang@physics.umass.edu}
\affiliation{Department of Physics,
University of Massachusetts,
Amherst, MA 01003-3720, USA}

\begin{abstract}
The densest packings of $N$ unit squares in a torus are studied using analytical methods as well as simulated annealing.  A rich array of dense packing solutions are found: density-one packings when $N$ is the sum of two square integers; a family of ``gapped bricklayer'' Bravais lattice solutions with density $N/(N+1)$; and some surprising non-Bravais lattice configurations, including lattices of holes as well as a configuration for $N=23$ in which not all squares share the same orientation.  The entropy of some of these configurations and the frequency and orientation of density-one solutions as $N \rightarrow \infty$ are discussed.

\end{abstract} 
\maketitle

\section{Introduction}

Understanding the dense packings of hard particles has yielded essential insights into the structure of materials \cite{Bernal1964,Zallen1983,Torquato2002,Chaikin2000}, granular media \cite{Torquato2002,Mehta1994}, number theory \cite{COHNa,Conway1999}, biology 
\cite{Gevertz2008,Purohit2003}, and computer science \cite{Johnson1974,Lodi2002}. This understanding has been hard won: hundreds of years can elapse between a conjecture and its proof. This is best exemplified by sphere packing, for which a proof Kepler's conjecture had not been found until 1998 \cite{HALESa}.

Recent experimental advances have allowed the development of (nearly) hard colloids that, for entropic reasons, manifest the densest sphere packing \cite{Pusey1986}. The attempt to obtain a deep understanding of liquid crystal mesophases has prompted study of the dense packing of anisotropic particles. This has led to recent explorations of the packing of ellipsoids \cite{Donev2004,Ras2011}, polyhedra \cite{ROAN,BAKER}, and polygons \cite{STROOBANTS}.

One of the simplest regular polygons one can pack in two dimensions is the square. On the plane, the densest packing is, in this case, trivial -- a square lattice of squares. Monte Carlo simulations of squares at finite pressure, however, have also found a tetratic phase \cite{Donev2006,Wojciechowski2004}, and experiments with hard colloidal squares have found, rather than the tetratic phase, a rhombic crystal having a different symmetry than the square \cite{Zhao2011}. Even the dense packing of a finite number of squares can be more complicated than naive considerations would indicate -- in fact, the problem of packing squares in another square has been shown to be NP-hard \cite{Leung1990}. Indeed, the densest known packings can be quite complex \cite{ERDOS1975,Friedman2002} when the number of squares is not a perfect square integer. Higher packing densities than that of a simple, square lattice with vacancies can be achieved through configurations in which some of the squares are rotated and shifted with respect to the square lattice \cite{Friedman2002}.

What is the effect of an external potential on packings of hard objects? One example of an external potential is fixed boundary conditions.  Considerable effort has been devoted to understanding the densest packings of squares and circles in various domains.  Another example with greater physical application is an external periodic potential.  If this potential is strong enough, the packing of the hard objects may be forced to be adopt the periodicity of the potential, and new packings are expected to arise.  A limiting case of an imposed periodic potential is periodic boundary conditions.  In this paper we explore the densest configurations of hard squares in a torus -- that is, inside a larger square with periodic boundary conditions.  Even with the additional translation symmetry afforded by packing squares in a torus rather than in a square, the resulting dense packings in the torus can still be far from simple. Our results may have experimental relevance for hard square colloidal particles in a periodic potential imposed either by a substrate or an optical lattice.

As in many other mathematical packing problems, the strategy here is to search for the smallest area torus that can accommodate a fixed number of squares $N$.  We use a combination of analytic and Monte Carlo simulated annealing techniques to accomplish this, and our results can be summarized as follows: we find that whenever $N$ can be expressed as the sum of two square integers -- $N=n_1^2+n_2^2$ -- the densest possible configuration is a density-one packing with squares arranged in rows that are oriented at an angle of tan$^{-1}(n_2/n_1)$ relative to the underlying torus.  For other $N$, we find a surprisingly rich collection of dense packing structures. For $N$ = 6,11,14, and 27, we believe that the densest possible packing is a commensurate Bravais Lattice packing with density $N/(N+1)$ and resembles a bricklayer pattern with periodic gaps.  For $N=$ 12,21,22 and 23, we find that the densest configurations are non-Bravais lattice packings, including both regular lattices of holes and of rotated squares. These results are summarized the Table within Section II.

In Section II, we present a summary analysis of the various structures we found for $N$ up to 27, including both commensurate Bravais lattice solutions and non-Bravais lattice solutions.  The packing motifs we found through analytic and numerical means are illustrated in this section via drawn figures as well as images generated by our numerical simulations. In Section III, we provide details of our numerical experiments for the hard square system. Section IV is a discussion of our results, including an analysis of the entropy of densest packings, and the rotational invariance of density-one packings as $N$ goes to infinity. 

\section{Analysis of Packings and Numerical Results}
\label{sec:analytics}

In this section we give an analytic treatment of square packings on a torus.  We first describe a set of solutions in which the squares lie on a Bravais lattice, and then turn to more complicated cases. The density-one solutions are optimal by construction, and all of the other solutions are conjectured to be optimal. The numerical results of Sec. \ref{sec:numerical} guided us to the conjectured solutions, and the fact that long simulated annealing runs consistently produced these solutions gives us some confidence that they are optimal.  Our conjectures for configurations containing $N \leq 27$ squares are summarized in the Table.

\subsection{Commensurate Bravais Lattice Solutions}
Here we consider a class of Bravais lattice configurations that includes all of the density-one packings and other solutions we have found for  $N \leq 27$.   In all of these packings, the squares are lined up in rows; and for the purposes of this analysis, we assume that these rows are aligned along the x-axis. Thus, one of the primitive vectors of the lattice of squares is $\ax=\hat{{\bf x}}$, and the second primitive vector is taken to have the form $\ay=c \hat{{\bf x}} + d \hat{{\bf y}}$, with $-1< c <1$ and  $|d|\geq 1$.  Note that the primitive vectors of the torus, $\Ax$ and $\Ay$, need not be aligned with the primitive vectors of the squares.     The requirement that the squares pack periodically on the torus is equivalent to saying that the lattice of squares is commensurate with the larger square lattice of the torus.  That is, there exist  integers $n_1$, $n_2$, $n_3$ and $n_4$ such that the torus primitive vectors $\Ax$ and $\Ay$ are given by
\begin{eqnarray} 
\label{eqn:Ana}
\Ax&=& n_1 \ax + n_2 \ay \nonumber \\ 
\Ay&=&n_3 \ax + n_4 \ay.
\end{eqnarray}
In addition, we require that the torus primitive vectors are of equal length,
\begin{equation}
\label{eqn:normal}
|\Ax|=|\Ay| ,
\end{equation}
and orthogonal,
\begin{equation}
\label{eqn:ortho}
\Ax \cdot \Ay = 0 .
\end{equation}

These conditions are uniquely solved by
\begin{eqnarray}\label{eq:gap}
c &=& - \frac{n_1 n_2 + n_3 n_4}{n_2^2 + n_4^2}\\
d &=& \frac{n_1 n_4 - n_2 n_3}{n_2^2 + n_4^2} \nonumber
\end{eqnarray}
The number of squares $N$ packed on the torus is the number of lattice points of the square lattice in a unit cell of the torus lattice
\begin{equation}
\label{eqn:N}
N = |n_1 n_4 - n_2 n_3|,
\end{equation}
and the areal density of the squares $\rho$ is given by
\begin{equation}
\label{ }
\rho=N/|\Ax \times \Ay| = 1/|d|.
\end{equation}

\subsubsection{Density-one packings}
\label{sec:densityOnePackings}
There are two classes of density-one packings.  The first is the perfect square packing, for which $c=0$, $d=1$, $n_1=n_4=\sqrt{N}$ and $n_2=n_3=0$.  This simple packing is shown in Fig. \ref{fig:N9} for the case of $N=9$.  Note that, on the torus, each of the $n_1$ rows (or columns, but not both) may be arbitrarily displaced relative to the other rows (columns) without disturbing the density of the packing or its periodicity; the perfect square packings thus have finite entropy.

\begin{figure}[h]
\label{fig:N9}
\scalebox{.5}{\includegraphics{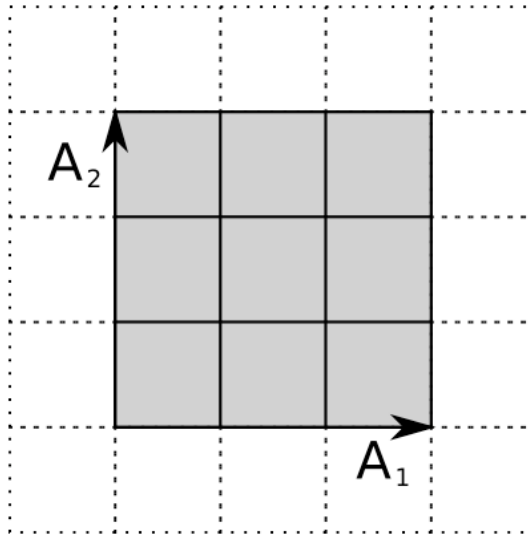}}
\caption{\label{fig:N9} An example of a perfect square packing with density one: $N=9$.}
\end{figure}

\begin{figure}[H]
\scalebox{.4}{\includegraphics{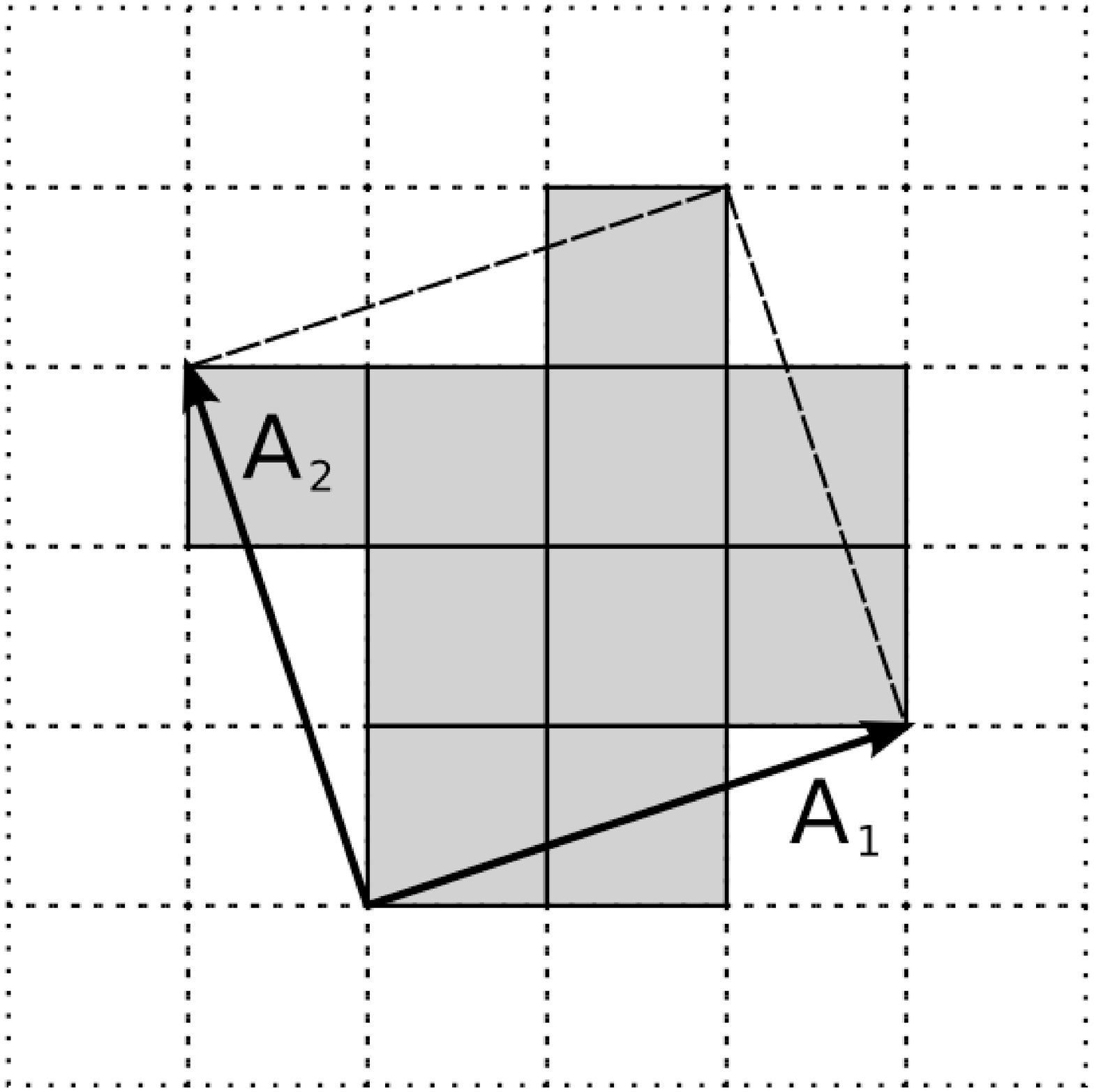}}
\caption{\label{fig:bravais} An example of a packing for which $N$ is equal to the sum of two squares: $N=10$; all such packings are density one.}
\end{figure}

There is a more general class of density-one packings, in which the lattice of squares may be tilted with respect to the primitive vectors of the torus. Setting $d=1$ and $c=0$ in Eq. (\ref{eq:gap}), we find $N=n_2^2 + n_4^2$, $n_1=-n_4$, and $n_2 = n_3$, the square lattice being oriented at an angle of $\tan^{-1}(n_2/n_1)$ relative to the torus lattice vectors.   Clearly, these density-one, tilted square lattice solutions are optimal for all $N$ that are sums of two square integers.  Note that the perfect square solution corresponds to the special case $n_2 = n_3 = 0$.  Fig. \ref{fig:bravais} shows the case $N=10$ ($n_1=3$ and $n_2=1$).

\begin{figure}[H]
\scalebox{.4}{\includegraphics{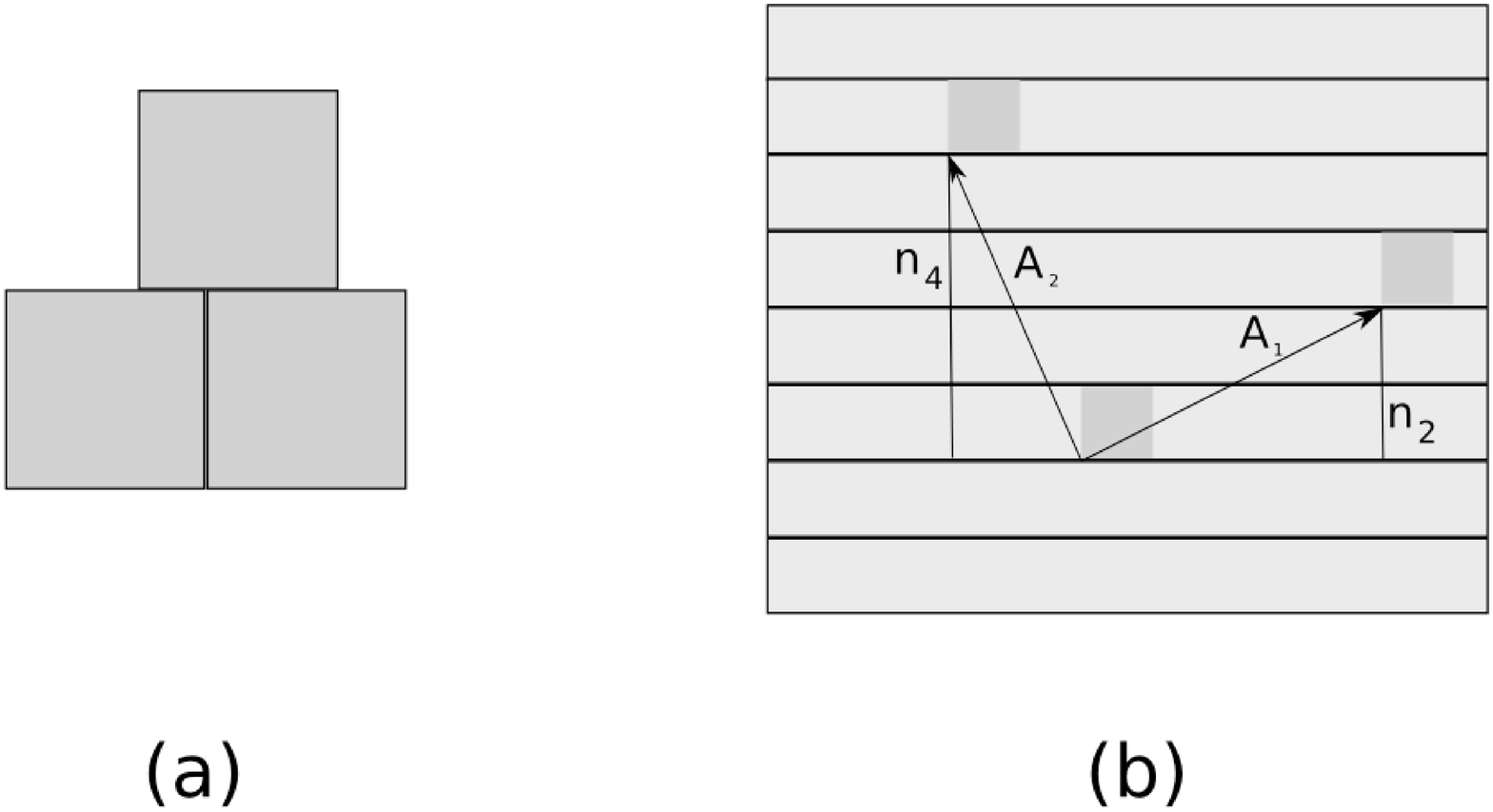}}
\caption{\label{fig:aligned} (a) Any three squares in mutual contact without gaps force two of the squares to define a row. (b) Diagram illustrating that $N$ must be equal to the sum of two squares for all density-one packings of squares in the torus (see discussion in section \ref{sec:densityOnePackings}).}
\end{figure}

We thus find that if N is the sum of two squares, there exists a density-one packing. The converse of this is also true: $N$ is a sum of two squares for all density-one packings of squares in the torus. To prove this, first note that every square in a density-one packing must have at least four other squares bordering it along a finite segment length, forcing all $N$ squares to share the same orientation.  Now consider three squares in mutual contact with each other -- a configuration that must exist if the packing has no gaps. Two of those squares must be aligned in a row, as shown in Fig. \ref{fig:aligned}. In order to eliminate gaps in the packing, these three squares define a set of rows that the entire packing must respect. Note that periodic boundary conditions allows us to draw the torus vectors so that they begin on the corner of a square and end on the corresponding corner of another square.  Thus, $n_2$ and $n_4$ are both integers. A right triangle can be constructed with $n_2$ as one side and $A_1$ as its hypotenuse; another right triangle can be drawn with $n_4$ as its base, and $A_2$ as its hypotenuse. (See Fig. \ref{fig:aligned}). These triangles are identical by inspection.  $|A_1|^2$ and $|A_2|^2$ are therfore each equal to $n_2^2 + n_4^2$, via the Pythogorean theorem.

\subsubsection{Lattice packings with vacancies}
\label{sec:vacancies}
\begin{figure}[H]
\scalebox{.48}{\includegraphics{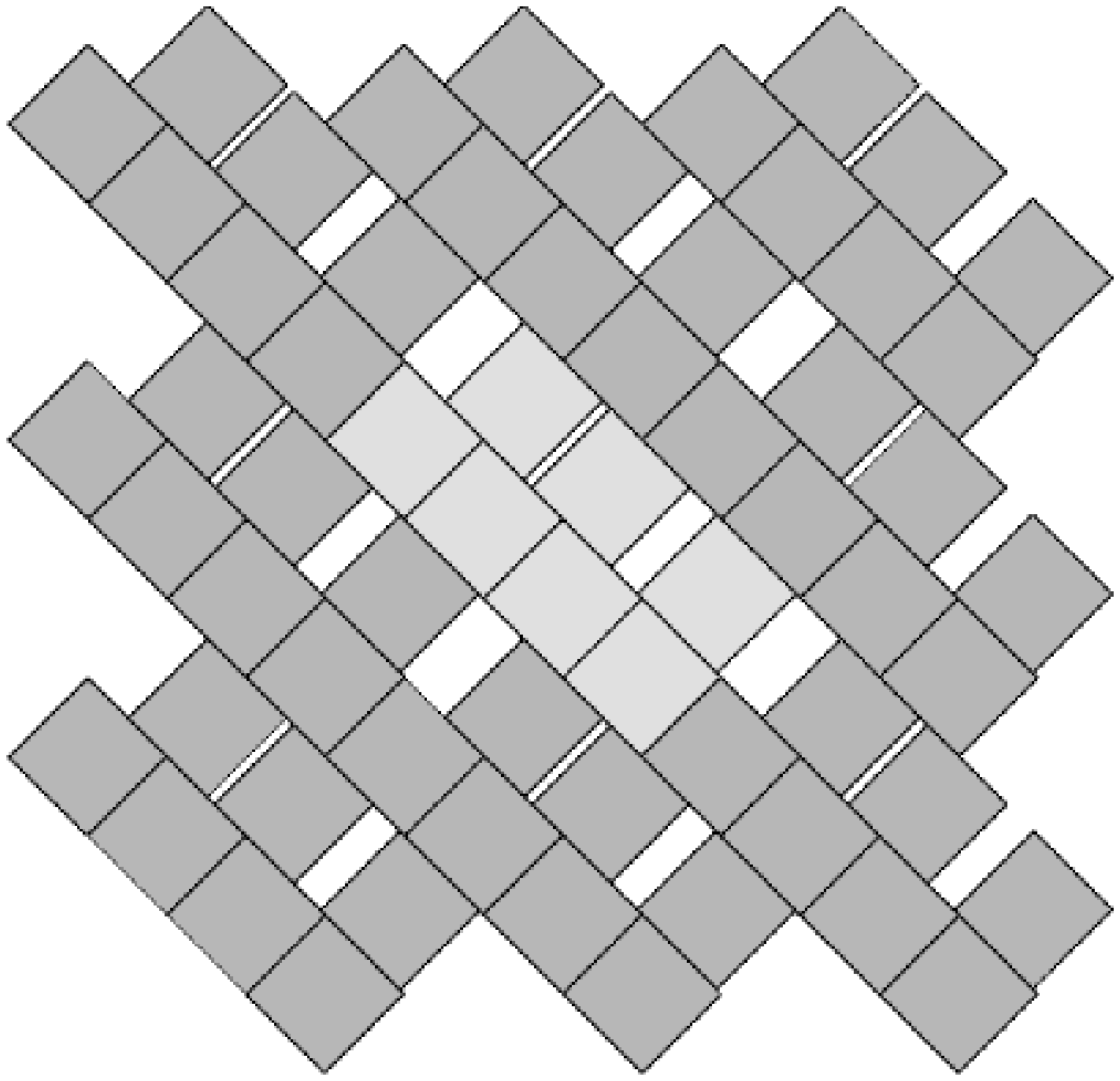}}
\caption{\label{fig:n7} An example of a packing with vacancies with the form $N=n_2^2 + n_4^2-k$. Here $N=7, k=1$ and $n_2=n_4=2$.}
\end{figure}

The simplest way to produce candidates for a densest packing for $N=n_2^2 + n_4^2-k$ is to remove $k$ squares from a density-one packing; indeed, our numerical results suggest that for several values of $N$, the densest packing is a density-one packing with one missing square.  This is indicated in the Comment column in the Table using the notation $n_1^2-1$ or $n_1^2+ n_2^2-1$, depending on whether they are generated by removing $1$ square from $N$ a square integer, or a sum of two square integers respectively.  As is demonstrated in Fig. \ref{fig:n7} for the case of $N=2^2+2^2-1=7$, such vacancies allow for continuous displacement of other squares within a row, leading to a finite entropy for such configurations.  Other examples include $N=3$ and $15$.

\subsubsection{Bricklayer packings with gaps}
\label{sec:bricklayer}
\begin{figure}[H]
\scalebox{.48}{\includegraphics{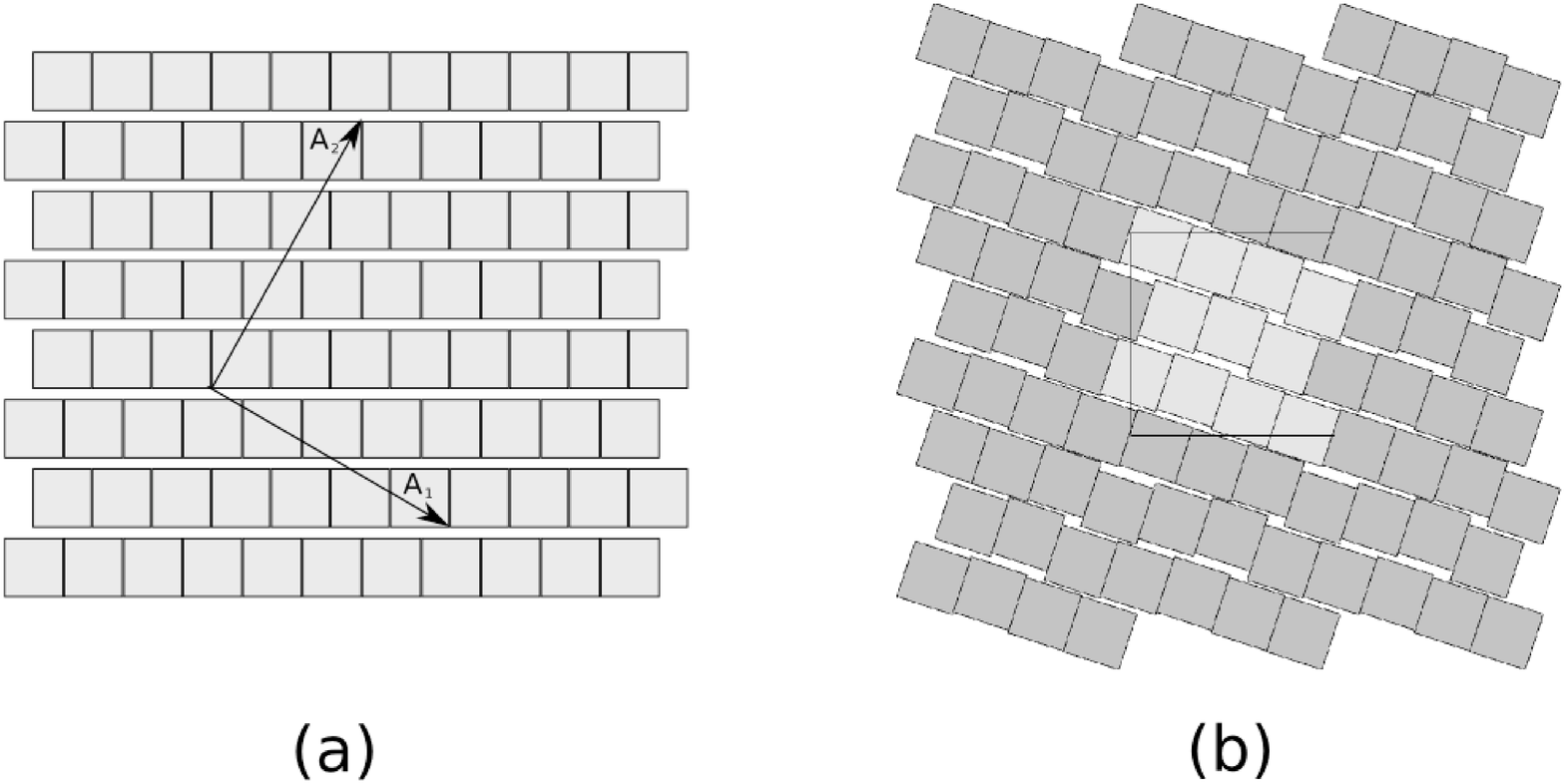}}
\caption{\label{fig:gb}Schematic (a) of a ``gapped bricklayer'' configuration, with density $\rho = (N-1)/N$. Results of simulated annealing for $N=11$ are shown in (b) ($n_1=3, n_2=1, n_3=-2, n_4=3$).  The finite entropy of this configuration is revealed by the displacements of the squares perpendicular to the rows.}
\end{figure}

Next we consider Bravais lattice solutions that have density less than one -- that is, packings with gap $d-1 > 0$.  Because these solutions have rows that are shifted relative to one another ($c\neq 0$), we call these ``gapped bricklayer configurations.''  An example is shown in Fig. \ref{fig:gb}(a).  Equations (\ref{eq:gap}) allow us to ennumerate all gapped bricklayer configurations. Since the packing density of these configurations is $(n_2^2 + n_4^2)/N$, the highest packing density we can find within this class of configurations with density less than unity must have the form $n_2^2 + n_4^2 = N-1$.  This requires that $N$ be one more than a sum of two squares. The first several are gapped bricklayer configurations are seen for $N=$ 2,3,6,11,14,18,26, and 27.  Based on the numerics we believe that for $N=6$, 11, 14 and 27, the gapped bricklayer packing is the densest possible packing. These are indicated in the Table with the abbreviation ``GB'' in the Comment column. Associated (non-unique) lattice vectors are shown in the rightmost columns of the Table for the GB packings.  There are also gapped bricklayer solutions for density $(N-2)/N$ when $N$ is two more than a sum of two squares, though we have not found any candidate densest packing solutions of this form for $N \leq 27$. Unlike the bravais lattice packings with density one, different rows of the gapped bricklayer solutions have a fixed shift given by $c = - (n_1 n_2 + n_3 n_4)/(n_2^2 + n_4^2)$, where the denominator is the closest sum of two squares below $N$. 

\subsection{Non-Bravais Lattice Packings}

Here, we consider special cases suggested by our numerical simulations that do not correspond to Bravais lattice packings.
Note: if the Comment column of the Table simply repeats a value of $N$, it indicates a special case for which the squares are not on a Bravais lattice and for which there is not an obvious pattern that can be extrapolated easily to optimal packings for higher values of $N$.

\subsubsection{Gapped bricklayer with domino bricks, $N=22$}
\begin{figure}[H]
\scalebox{.7}{\includegraphics{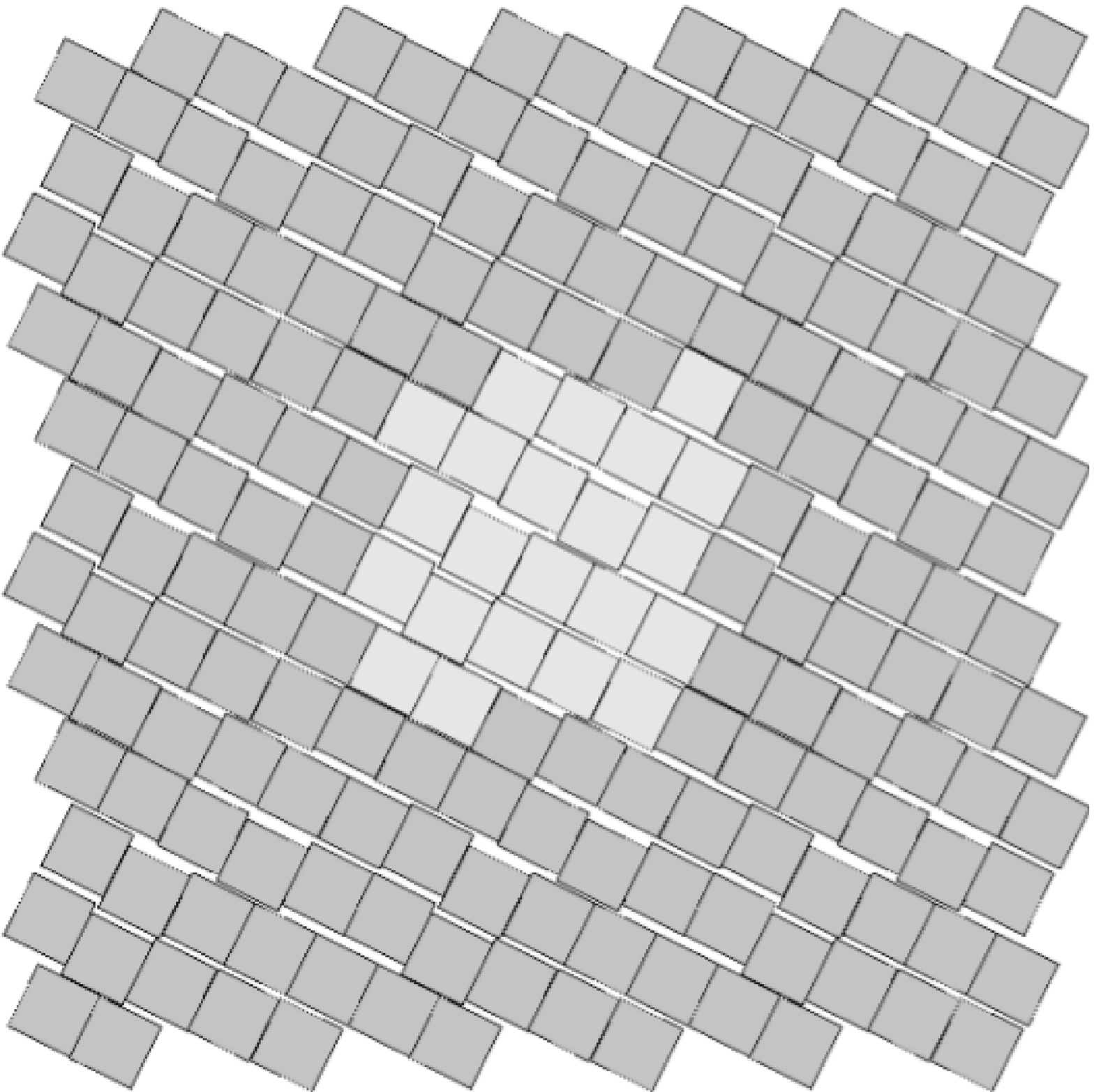}}
\caption{\label{fig:n22}Simulation results for $N=22$: the conjectured best packing is a ``gapped bricklayer with domino bricks''.}
\end{figure}

The conjectured best packing for $N=22$ is shown in Fig. \ref{fig:n22}. This packing is in fact a gapped bricklayer configuration, except that the unit cell or brick is composed of two squares stacked in the $\hat{{\bf y}}$ direction (the direction perpendicular to the rows).  The configuration is otherwise identical to the $N=11$ gapped bricklayer and has density $\rho= 10/11$.

\subsubsection{Lattice of $\frac{1}{2}\times\frac{1}{2}$ holes, $N=12$ and $23$}
\begin{figure}[H]
\scalebox{.7}{\includegraphics{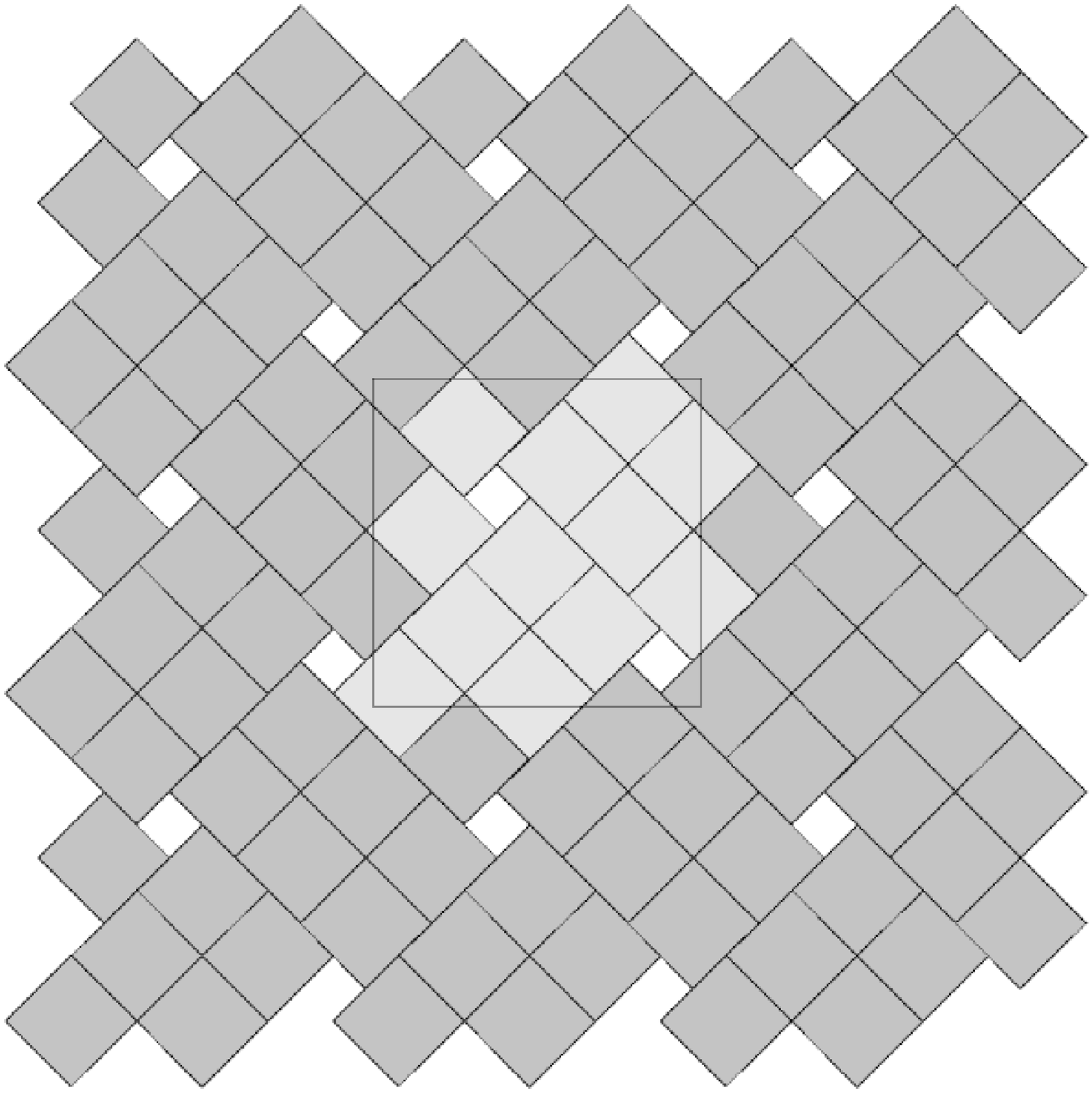}}
\caption{\label{fig:n12}Simulation results for $N=12$: a lattice of $\frac{1}{2} \times \frac{1}{2}$ holes.}
\end{figure}

\begin{figure}[H]
\scalebox{.65}{\includegraphics{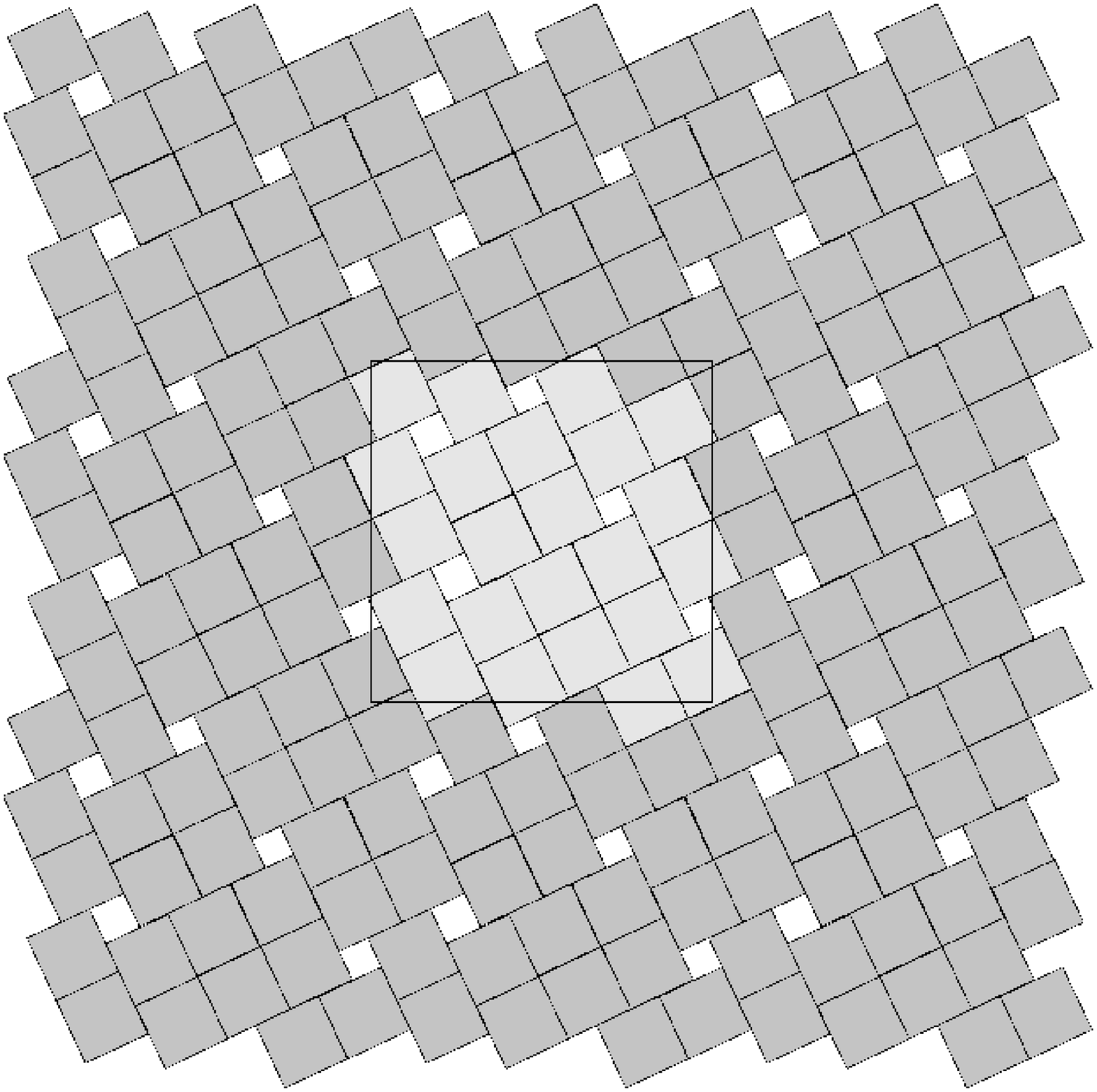}}
\caption{\label{fig:n23}Simulation results for $N=23$: a lattice of $\frac{1}{2} \times \frac{1}{2}$ holes.}
\end{figure}

The conjectured best configurations for $N=12$ and 23 are shown in Figs.\ \ref{fig:n12} and \ref{fig:n23}, respectively.  In both cases the motif can be described as a lattice of $\frac{1}{2}\times\frac{1}{2}$ holes.  The torus lattice vectors, Eq.\ \ref{eqn:Ana} with $\ax=\hat{{\bf x}}$ and $\ay=\hat{{\bf y}}$, for $N=12$ are described by $n_1=n_2=-n_3=n_4=5/2$ and the torus lattice vectors for $N=23$ are described by $n_1=n_4=9/2$, and $-n_2=n_3=2$.  It is straightforward to verify that these motifs are in fact packings on the torus and have the density $N/(n_2^2 + n_4^2)=N/(N+k/4)$ where $k$ is the number of holes in the unit cell.  For $N=12$, evidently $k=2$ and for $N=23$, $k=5$. 

\subsubsection{Lattice of skew squares embedded in a square lattice, $N=21$}
\begin{figure}[H]
\scalebox{.5}{\includegraphics{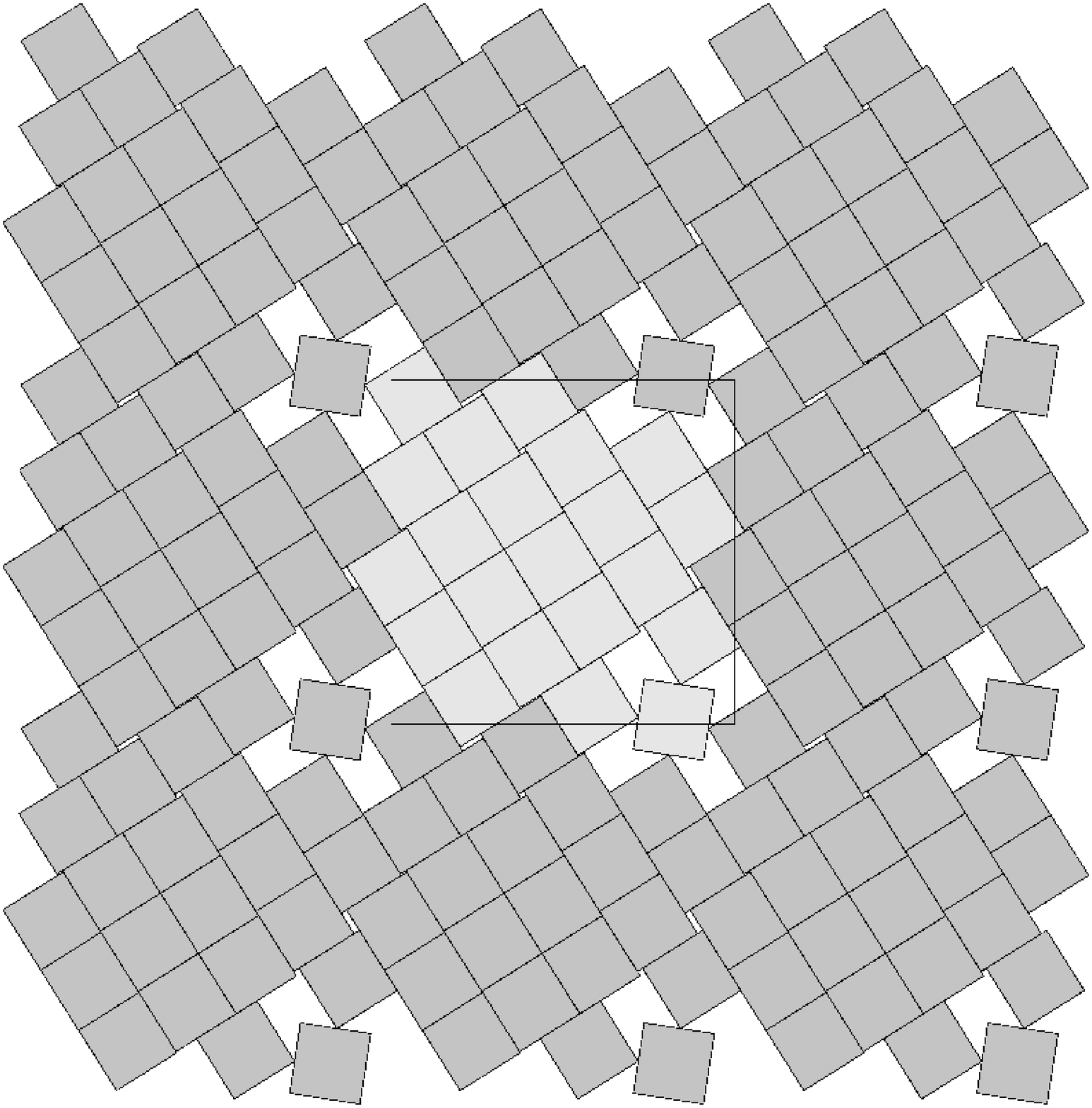}}
\caption{\label{fig:n21}Simulation results for $N=21$: a lattice of skew squares embedded in a square lattice.  The $5$-square pattern (which includes a skew square in its center) is the proved best packing of 5 squares in a square \cite{Friedman2002}. Note that the simulation results have not yet converged to the conjectured best packing.} \end{figure}

The conjectured densest packing for  $N=21$, shown in Fig.\ \ref{fig:n21}, does not follow any of the motifs described heretofore. The unit cell consists of a $4 \times 4$ square with motif of 5 squares attached to its side.  This 5-square pattern is also the best packing of 5 squares in a square \cite{Friedman2002}.  A simple calculation yields the density, $\rho= 21/(4^2+(2+1/\sqrt{2})^2)$. This packing has one square per unit cell tilted at 45$^{\circ}$ relative to all other squares.  This is the only example that we found for which not all of the squares in the motif are oriented in the same way. It was also the most difficult configuration for our simulated annealing algorithm to find. Figure \ref{fig:n21} shows a typical simulation result, which clearly has not yet fully converged.

\subsection{Table of Results}

To summarize: the perfect square, sum of two squares and gapped bricklayer configurations cover most of the case we have found for $N \leq 27$.  The Table gives densest packing configurations (if $\rho=1$) and conjectured densest packing configurations (if $\rho<1$), for each value of  $N$ less than 28.  The column ``$\rho$'' is the density of the configuration. The Comment column describes the type of lattice.   For example, $3^2$ indicates a perfect square and $5^2-1$ indicates a perfect square with one square missing.  Similarly $3^2+1^2$ refers to the sum of two squares and ``GB'' stands for ``gapped bricklayer.''  If a single number is in the Comment column it refers to one of the special cases discussed above.  The four columns ``$n_1$, $n_2$, $n_3$, $n_4$'' are shown if the configuration of squares is itself a Bravais lattice; these integers are the coefficients of the lattice vectors of the torus, in terms of the lattice vectors of the squares as defined in Eq. (\ref{eqn:Ana}).  Only those columns needed to specify the lattice are filled in.

\vspace{.3in}
\begin{table}
\label{table}
\caption{Exact and conjectured densest packed configurations of squares on a torus.  Refer to the text for the meaning of the columns.}
  \begin{tabular}{|c | c|c | c| c | r | c | c | c |}
\hline
  $N$ &$\rho$&Comment& $n_1$ & $n_2$ & $n_3$ & $n_4$  \\ \hline \hline
   1 &1&  $1^2$ & 1&  &   &    \\ \hline 
   2 &1&  $1^2+1^2$ &1& 1 &  &  \\ \hline
   3 &$3/4=0.75$&  $2^2 -1$ &2&  & &   \\ \hline
   4 &1&  $2^2$ &2& & &   \\ \hline
   5 &1&  $2^2+1^2$ &2& 1 &    & \\ \hline
   6 &$5/6=0.8\bar{3}$& GB &2& -1 &2 & 2   \\ \hline
   7 &$7/8=0.875$& $2^2+2^2 -1$ &2& 2 & &     \\ \hline
   8 &1& $2^2+2^2 $ &2& 2 & &     \\ \hline
   9 &1&  $3^2$ &3& & &  \\ \hline
   10 &1&  $3^2+1^2$ &3& 1 &    & \\ \hline
   11 &$10/11=0.\overline{90}$& GB &3& 1 &-2 & 3   \\ \hline
   12 &$24/25=0.96$& 12 && & &    \\ \hline
   13 &1&  $3^2+2^2$ &3& 2 &    & \\ \hline
   14 &$13/14=0.9\overline{285714}$& GB &4& -2 &1 & 3   \\ \hline
   15 &$15/16=0.9375$&  $4^2 -1$ &4&  & &   \\ \hline
   16 &1&  $4^2$ &4& & &   \\ \hline
   17 &1&  $4^2+1^2$ &4& 1 &    & \\ \hline
   18 &1&  $3^2+3^2$ &3& 3 &    & \\ \hline
   19 &$19/20=0.95$&  $4^2 +2^2 -1$ &4& 2 & &   \\ \hline
   20 &1&  $4^2+2^2$ &4& 2 &    & \\ \hline
   21 &$21/(4^2+(2+1/\sqrt{2})^2)=0.900189 \ldots$&  $21$ & &  &   &    \\ \hline 
   22 &$10/11=0.\overline{90}$& 22 &3& 1 &-2 & 3  \\ \hline
   23 &$92/97=0.94845 \ldots$&  $23$ & &  &   &    \\ \hline
   24 &$24/25=0.96$&  $5^2 -1$ &5&  & &   \\ \hline
   25 &$1$&  $5^2$ &5&  & &   \\ \hline
   26 &1&  $5^2+1^2$ &5& 1 &    & \\ \hline
   27 &$26/27=0.\overline{962}$& GB &5&1 &-2 & 5  \\ \hline

\hline
\end{tabular}
\end{table}

\section{Numerical Methods}
\label{sec:numerical}

For all $N \leq 27$ squares on the flat torus, we searched for densest packings of $N$ squares on the torus via Monte Carlo simulations in the NPT ensemble.  Our approach was to employ a simulated annealing (SA) algorithm in which the system was taken from an initial, low-pressure, easy-to-equilibrate state to a final, high-pressure state, via an annealing schedule consisting of a series of steps in inverse pressure.  Between each pressure increase a Metropolis algorithm appropriate to the hard square NPT ensemble was used to equilibrate the system.  Although SA quickly falls out of equilibrium at higher pressures as the energy landscape becomes rough, it appears to be an an effective algorithm for finding ground states of the system.

The equilibration procedure we used in our simulated annealing algorithm was a Metropolis procedure consisting of three types of Monte Carlo moves:  translations and rotations of individual squares, and changes in the volume of the entire system.  At each step of the equilibration procedure, a square is selected at random; then, one of the three types of moves is selected at random, with probabilities .495, .495, and .01, for translation, rotation, and volume change, respectively.  Once a square and a move type is selected, the move is attempted.  If the move results in any overlaps among the $N$ squares, the move is rejected.  If it does not, then for translations or rotations, the move is accepted; for volume change $dV$, the move is accepted with probability $p_{acc}=\min[1,\exp(-\beta P dV)]$.  In practice, rather than changing the volume of the entire system, the periodic box in the simulations was kept at a constant size, and the sizes of the individual squares were all rescaled, in order to achieve the desired new volume. The equilibration procedure consists of $s$ such Monte Carlo steps; in our simluations $s$ was typically between 200 and 400 steps.  

We repeated this simulation 1,000 times and reported the highest density found among these runs.  In order to determine the highest-density packing for $N$ that did not correspond to a perfect square or a sum of squares, more extensive runs were conducted -- in some cases, as long as 72 hours on a 2GHz processor.  The fact that significantly different initial configurations as well as different initial random seeds generated the same final, high-density configurations signalled that a good candidate for a densest packing of the system had been found.  For all simulations, the pressure was initially set to $\beta P=.01$, and was increased via constant steps in inverse pressure until a maximum pressure of $\beta P_{max}=3000$ was reached. (During subsequent explorations of the phase behavior of the system, this pressure was later deemed excessive, but nevertheless produced reasonable results for the purposes of determing the ground state of the system.) All simulations were begun at an initial areal density $\rho$ of 0.1, with a square array of unit squares. Before each equilibration procedure began, a trial run was conducted in which the maximum value of translations, rotations, and volume changes were independently optimized in order to achieve an acceptance ratio for each of $0.4$. 

The majority of the computational work during the equilibration procedure consisted in checking for square overlaps.  For this, we relied on a fast algorithm for detecting polygon overlaps by Alan Murta \cite{Murta}, and an associated Python wrapper by Joerg R\"adler \cite{Radler}. Configurations were visualized using the VPython library \cite{Scherer2000}.

\section{Discussion}
\label{sec:discussion}

In this paper, we have presented an analysis of the densest packing solutions for $N$ unit squares in the torus.  For $N \leq 27$, the majority of these packings are Bravais lattice solutions, manifested either as the ``sum of two square integers'', or ``gapped bricklayer configurations'' described above. A few were non-Bravais lattice solutions, such as those $N=21$ and $N=23$. In this section, we discuss the frequency and entropy of the various types of packings we found, and pose some questions for further study.

We showed in Section \ref{sec:densityOnePackings} that density-one packings are only possible for those values of $N$ that are expressible as the sum of two squares. 
Though it appears that density-one packings are relatively common from the Table, in fact it is known that the frequency of numbers that are equal to the sum of two square integers scales as $1/\sqrt{\ln N}$ for large $N$ \cite{Berndt1993,Landau1909}.
Thus the frequency of density-one packings also vanishes with increasing $N$.

Despite the relative scarcity of density-one packings, we argue that the packing density approaches one as $N \rightarrow \infty$. Suppose $M$ is a sum of two square integers. Construct a new packing for $N = M-k$ by removing $k$ squares. The resulting packing density gives a lower bound of $\rho=N/(N+k)$. Given a number of squares, $N$, however, determining $k$ requires knowledge of the nearest density-one packing larger than $N$. For sufficiently large $N$, we can estimate that the distance to the next density-one packing is of order $\sqrt{\ln N}$ larger than $N$. Therefore, an estimated lower bound on the packing density of $N$ squares is $1-\sqrt{\ln N}/N$ for large $N$, which yields the result that the density approaches one asymptotically. This argument does not take into account fluctuations in the spacing of sums of two squares, and it would be interesting to find a mathematically rigorous asymptotic lower bound on the packing density.

The main contributions to the entropy of the various Bravais and non-Bravais lattice configurations for $N \leq 27$ are readily assessed. 
 For a Bravais lattice packing, there are no non-trivial continuous symmetries and thus there is no entropy.  However, one can construct density-one packings from the Bravais lattice packings by shifting rows relative to one another.  For example, if $N$ is a perfect square then each row can be arbitrarily shifted and, ignoring overall shifts of the lattice, the entropy is proportional  ($\sqrt{N}-1$).  Note that either rows or columns may be shifted but not both for density-one packings.  More generally, rows are free to shift unless the contraints of periodicity forbid it.  Periodicity forbids two rows from shifting relative to one another if some linear combination of torus lattice vectors connects the rows.  Suppose that rows are aligned in the $\hat{\bf x}$-direction.  Then, two rows are constrained to their relative positions in the Bravais lattice configuration if there are integers $a$ and $b$ such that $a \mathbf{A}_1 + b \mathbf{A}_2$ connects the two rows. Thus, rows separated by $a n_2 + b n_4$ are locked (see Eq.\ \ref{eqn:Ana} and Fig.\  \ref{fig:aligned} (b)).
Bezout's Lemma \cite{Jones1998} states that this integer linear combination can be made equal to, but not smaller than, $g$, where $g$ is the greatest common divisor of $n_2$ and $n_4$ (assuming that both $n_2$ and $n_4$ are nonzero). Thus, the set of rows can be divided into $g$ groups, each of which can be arbitrarily shifted, and the entropy is proportional to $g-1$.
Note that if $|n_2|$ and $|n_4|$ are mutually prime, all rows are locked and this contribution to the entropy of the configuration vanishes. There are two other sources of entropy for packing related to Bravais lattice packings. The gapped bricklayer configurations (Section \ref{sec:bricklayer}) allow squares within a row to shift perpendicular to the row axis (see Figures \ref{fig:gb} and \ref{fig:n22}), and this ``poor workmanship'' contribution to the entropy will be roughly proportional to the free volume of the configuration.  The density-one configurations with vacancies ($n_2^2+n_4^2-k$), discussed in Section \ref{sec:vacancies}, also possess finite entropy, since the hole(s) created by the $k$ missing squares may be moved throughout the lattice, or split along a row; and for $k>1$, holes can appear in different rows.
In contrast, the unusual, non-Bravais lattice packings of the  $N=12$ and $N=23$ exhibit no entropy.

The above entropies have implications for which configurations are most likely to be seen at finite pressure. For example, $N=25$ admits of two classes of configuration:  a density-one packing with rows aligned along the torus ($N=5^2$) or with rows oriented at an angle with respect to the torus ($N=3^2+4^2$).  As we have seen, the $N=5^2$ has four rows that are free to slide while $N=3^2 + 4^2$ is a locked configuration with no entropy; this implies that the $N=5^2$ packings will be much more likely to appear at finite pressure.

Unlike squares packed into a square boundary, squares packed on a torus maintain rotational invariance in the thermodynamic limit.  This can be seen as follows: one can see in Figure \ref{fig:bravais} that any $N=n_1^2+n_2^2$ packing (which are, as seen above, the only possible packings with density one) will orient the square lattice at an angle of $\tan^{-1}(n_{2,i}/n_{1,i})$ relative to the torus lattice vectors.  To take the thermodynamic limit with a particular square lattice orientation $\Theta$ with respect to the underlying torus, it is sufficient to choose a particular subsequence of integers $N_i=n_{1,i}^2+n_{2,i}^2$ such that 
$\tan^{-1}(n_2/n_1) \rightarrow \Theta$.
The thermodynamic limit of density-one packings on the torus thus preserves rotational symmetry.

Our study of the densest packings of $N$ unit squares in a torus has yielded definitive results for cases in which $N$ is the sum of two square integers or is a perfect square, and strong conjectures for other values of $N \le 27$.  This work raises many interesting questions: How common are densest packings that have squares with different orientations, such as occurs for $N=21$?  Which motifs, if any, dominate for large $N$?  Are the $1/2 \times 1/2$ motifs of $N=12$ and $N=23$ exhibited for other $N$? What role do these dense packing configurations play in the thermodynamic phases observed in finite-temperature simulations and experiments with square colloids?

\begin{acknowledgements}
We would like to thank James Hanna and Narayanan Menon for useful discussions of some of the issues addressed in this paper.  J.M. acknowledges support from NSF grant DMR-0907235. C.D.S. acknowledges support from NSF grant DMR-0846582, and was partially supported by the NSF-funded Center for Hierarchical Manufacturing, CMMI-1025020. D.B. is grateful for use of the Hoffman2 computing cluster at UCLA.
\end{acknowledgements}

\end{document}